\documentclass[english,twocolumn,preprintnumbers,amsmath,amssymb,groupedaddress]{revtex4}
\usepackage{verbatim}
\usepackage{graphicx}
\usepackage{amssymb}
\usepackage{babel}
\usepackage{float}
\usepackage{amsmath}
\usepackage{amssymb}
\usepackage{graphicx}
\usepackage{verbatim}
\usepackage[unicode=true,pdfusetitle,
 bookmarks=true,bookmarksnumbered=false,bookmarksopen=false,
 breaklinks=false,pdfborder={0 0 0},backref=false,colorlinks=false]
 {hyperref}

\makeatother

\begin{document}

\preprint{XXX}

\title{A model independent approach towards resource count and precision limits in a general measurement}

\author{H. M. Bharath}
\email{bharath.hm07@gmail.com}
\author{Saikat Ghosh}
\email{gsaikat@iitk.ac.in}
\affiliation{
Department of Physics,
Indian Institute of Technology, Kanpur,
208016, India}

\date{\today}

\begin{abstract}

A formulation towards quantifying resource count used in a measurement, that is independent of the model of the measurement dynamics(Quantum/Classical), is considered.
For any general measurement with $(M+1)$ discrete outcomes, it is found that there is a unique probability distribution that minimizes the measurement error, with the error scaling as $1/M$. For a measurement with a finite resource(R), this absolute bound implies the resource count to be equal to the possible outcomes i.e. $R=M$. This formulation therefore provides a model independent route towards estimating resource count used in any general measurement scheme.

\end{abstract}


\maketitle
\maketitle
\section{Introduction:}
\par Measuring a physical parameter precisely has been one of the most fundamental pursuits in science. 
Improved accuracy in estimating the value of an unknown parameter has almost always lead to new physical insights together with an access 
to new and unexplored phenomena \cite{Gen-1,Gen-2,Gen-3}. One therefore wonders if there is a fundamental limitation on how precisely a
parameter can be measured. For a specific experiment, the answer depends on the amount of $\it{resources}$ used for the measurement:
quantifying limits on precision thereby gets redefined to quantifying the amount of resources $(R)$ used. 

\par Resources have been quantified earlier in varied ways. For every such way, there is a corresponding error bound. The simplest of these equates the resource count($R$) to the total count of the number of times an experiment is repeated, or the number of probe particles used($N$). The corresponding bound is known to scale as $(\frac{1}{\sqrt{N}})$ for models using classical strategies (following the central-limit theorem).  However, it has been found that quantum mechanically correlated probe states can significantly improve the precision, bounded by the Heisenberg Limit(HL) of $(\frac{1}{N})$\cite{Gen-1,Gen-2,Gen-3,QMetro-1, QMetro-2, QMetro-3}. In a more general approach, resource has been quantified as the count of number of times the measurement system is sampled\cite{QMetro-1, Resource-1, Resource-2, Resource-3} or  via generalized versions of uncertainty relations\cite {CRBound-3, Alt-bounds-1, Alt-bounds-2, Alt-bounds-3}.

\par All of these approaches of quantifying resources in a model depend on specificities of the dynamics considered in the measurement process. However, independent of the model, information of precision is contained in its output, which is a set of numbers(the possible outcomes) with associated probabilities. It is therefore expected that, a general quantification of resource that uses only the set of possible outcomes and the associated probabilities, can be developed. This characterization would be independent of the specific dynamical model. 

\par In this paper, we explore this approach towards quantifying resource count($R$). The error in a given measurement depends only on the output probability distribution, and varies with different distributions. One can therefore ask: Is there a distribution, for a given set of outcomes, that minimizes the error? 

\begin{figure}[H]
 \includegraphics[scale=0.57]{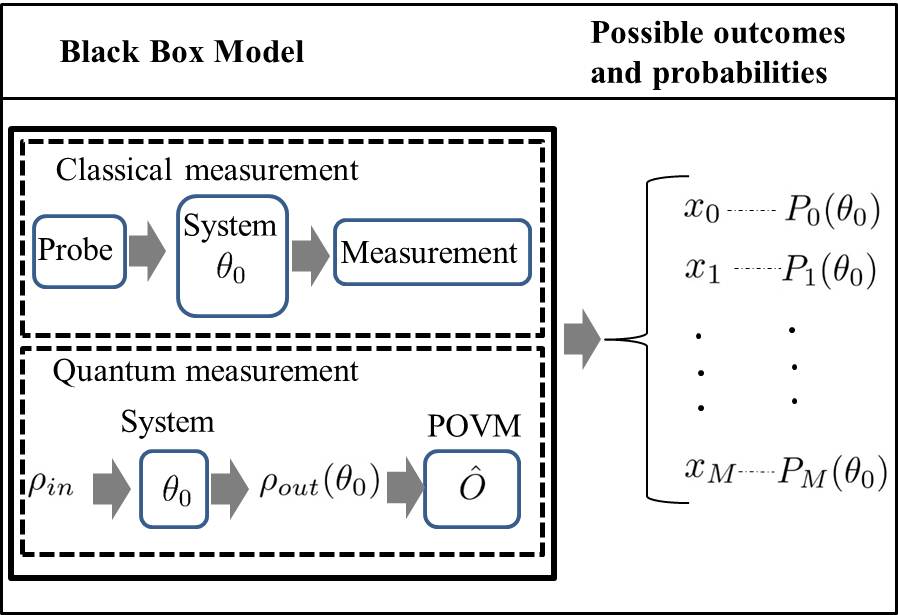}

\caption{ A cartoon of a general measurement model: In a classical measurement, the probe, after interacting with the system whose parameter $\theta_0$ is to be estimated, is characterized with a classical distribution $\{P_0(\theta_0),P_1(\theta_0)\cdots P_M(\theta_0)\}$ over the possible outcomes $\{x_0,x_1\cdots x_M\}$.
 In a quantum measurement, starting with a state $\rho_{in}$, the probe evolves to a state $\rho_{out}(\theta_0)$. For an appropriate POVM, the outcomes $\{x_0\cdots x_M\}$ are the eigen-values of the observable $\hat{O}$, being measured. This observable has a spectral decomposition $\hat{O}= x_0\hat{\Pi}_0+\cdots +x_M\hat{\Pi}_M$, where $\hat{\Pi}_k$ is the projection in to the eigen-space of $\hat{O}$ with eigen-value $x_k$. The probabilities for this case are given by  $P_k(\theta_0)=Tr\{\hat{\Pi}_k \rho_{out}(\theta_0)\}$. If viewed as a black-box, both models output a set of  possible outcomes $\{x_0,x_1 \cdots x_M\}$  with corresponding distributions.
}
\end{figure}

\par Here, we explicitly construct such a distribution and show that it is unique. The corresponding error is found to scale inversely with the number of outcomes(M). Since no particular dynamics(classical or quantum)is assumed, any measurement scheme that yields a finite, discrete set of outcomes will follow this error bound. Interestingly, it turns out to be impossible to design an estimation strategy that can achieve this bound, exactly. Furthermore, we propose an ansatz, motivated by the scaling of the error, characterizing the count of resources used in the estimation $(R)$, as the number of possible outcomes($M$).

\par For a fixed distribution, the error depends on the information contained in the distribution. Fisher information precisely quantifies this for any arbitrary output probability distribution while a lower bound for the error of an unbiased estimator, corresponding to the given distribution is set by the Cramer-Rao inequality\cite{CRBound-1,CRBound-2,CRBound-3}. Here, we show that the unique distribution that minimizes the error, saturates the Cramer-Rao inequality. Also, we show that this distribution maximizes the Fisher information.

\section{Problem Statement:}

We limit ourselves to an analysis of a general measurement model for estimating a single parameter with a natural(or true) value $\theta_0$, that has a discrete set of outcomes $\{x_0,x_1\cdots x_M\}$ with corresponding probabilities $\{P_0(\theta_0),P_1(\theta_0)\cdots P_M(\theta_0)\}$. The number of outcomes here is $M+1$. The expected value is then equal to the natural value: $ {\sum\limits_{k=1}}^M x_k P_k(\theta_0)=\theta_0$ while, $P_k(\theta_0)$ is the probability of obtaining $x_k$ as the outcome or estimated value from a measurement.(Fig. 1)

For an estimate of the parameter, $\theta_{est}=x_r$, one then needs to infer the natural value $\theta_0$, using the constraint $|\theta_0-\theta_{est}|\leq \sigma(\theta_0)$. Here, $\sigma(\theta_0)=\sqrt{ \sum_kx_k^2P_k(\theta_0)-\theta_0^2}$ is the root-mean-square-error(RMSE) of the distribution\cite{Variance-1}. The range of values of $\theta_0$ which satisfy the above inequality with $\theta_{est}=x_r$ (the obtained value), is the inferred range of the natural value. The width of this range is the precision of the measurement.
 
  \textbf{Example:}
Let us consider a measurement with two possible outcomes, say, $\pm1$. The parameter to be estimated, $\theta_0$ is in the range $[-1,+1]$. The associated probabilities at the end of the system-probe interaction are  $\frac{1\pm\theta_0}{2}$, so that the expected value is $\theta_0$. This form of dependence of the probabilities on $\theta_0$ is directed by the dynamics of interaction. The value of $\theta_0$ is an unknown of the system. The probe \textit{picks up} this value during its interaction with the system. It is to be estimated after a measurement. 

Suppose that we repeat this experiment $N$ times. The possible values of the average are $\{-1, -1+\frac{2}{N} \cdots 1-\frac{2}{N}, 1\}$. These are the $N+1$ possible outcomes of the overall experiment, i.e, this is the set: $\{x_0\cdots x_M\}$.
The associated probability distribution is a binomial: 
\[
P_k(\theta_0)=\left( \begin{array}{c} N \\ k \end{array}\right)\bigg(\frac{1-\theta_0}{2}\bigg)^k\bigg(\frac{1+\theta_0}{2}\bigg)^{N-k}.
\]

The only unknown is $\theta_0$, which is the peak of this distribution. Where ever the peak is, a measurement result will have to be within the peak's width: i.e., wherever $\theta_0$ is, the estimate obtained from a measurement, $\theta_{est}$ will be within the range $[\theta_0-\sigma(\theta_0),\theta_0+\sigma(\theta_0)]$. Therefore, this estimate can be used to infer the possible values of $\theta_0$.

Now, suppose that we obtained an estimate, $\theta_{est}$ after the measurement. This will be one of the $N+1$ possible outcomes: $\{-1,-1+\frac{2}{N}\cdots 1-\frac{2}{N},1\}$. We can then conclude that the value of $\theta_0$ is such that, it includes the value $\theta_{est}$ in it's width, i.e., $\theta_0$ satisfies the inequality $|\theta_0-\theta_{est}|\leq \sigma(\theta_0)$. The standard error $\sigma(\theta_0)$ is in this case given by $\sigma(\theta_0)=\sqrt{\frac{1-\theta_0^2}{N}}$.

Solving for the range of $\theta_0$ that satisfies the above inequality, we can show that asymptotically, this range is $\theta_0 \in [\theta_{est}-\frac{1}{\sqrt{N}}, \theta_{est}+\frac{1}{\sqrt{N}}]$. Whence, the error in this estimation is $\frac{1}{\sqrt{N}}$.  

This is the error for a binomial distribution. In general, we can design an experiment that assigns any other probability distribution $P_k(\theta_0)$, that correlates the $N$ repetitions. Using the corresponding standard error $\sigma(\theta_0)$, we can obtain the range of $\theta_0$ that can be inferred from an estimate $\theta_{est}$. This error may be smaller than $\frac{1}{\sqrt{N}}$, depending on the probability distribution.

\par The question we ask here is, for a general set of outcomes $\{x_0\cdots x_M\}$ what is the associated distribution $P_k(\theta_0)$ that minimizes the error? Further, what is the value of this minimum error?. Such a distribution, if unique, would then provide a standard for comparing a wide range of measurement models with varied nature of resources \cite{NonLin-1,NonLin-2,NonLin-3}. This distribution will be referred to hereafter as the Minimum-Error-Distribution(MED).

\section{Minimum Error Distribution:} To construct such a distribution with the least error, one can arrange the discrete $M+1$ outcomes of the model, $\{x_0,\cdots x_M \}$ in ascending order, such that the interval $[x_0,x_M]$ is divided in to $M$ bins: $[x_0,x_1], [x_1,x_2]\cdots [x_{M-1},x_M]$. The expected value $\theta_0$ falls in to one such bin, say $[x_r,x_{r+1}]$. One can then construct the MED with $P^{MED}_k(\theta_0)=0$ for every $k$, sparing the two special points $r$ and $r+1$(Fig.2(C)).

\par The probabilities $P^{MED}_{r}(\theta_0)$ and $P^{MED}_{r+1}(\theta_0)$ are trivially assigned from the two constraints of the problem: (1)$x_rP^{MED}_r(\theta_0)+x_{r+1}P^{MED}_{r+1}(\theta_0)=\theta_0$ is the expected value and (2) the distribution is normalized:$ P^{MED}_r(\theta_0)+P^{MED}_{r+1}(\theta_0)=1$.
Formally, the MED can be stated as:
\begin{equation}
P^{MED}_k(\theta_0)=\begin{cases}
\frac{\theta_0-x_{k-1}}{x_{k}-x_{k-1}} & x_{k-1}\leq \theta_0 \leq x_{k}\\
\frac{x_{k+1}-\theta_0}{x_{k+1}-x_{k}} & x_{k}\leq \theta_0\leq x_{k+1}\\
0 & \theta_0>x_{k+1} \quad or \quad \theta_0<x_{k-1}
\end{cases}
\end{equation}

Using Cauchy-Schwartz inequality, it is straightforward to show that this distribution has a lower RMSE than any other arbitrary distribution $\{q_0,q_1\cdots q_M\}$ with the same expected value $\theta_0$ (Fig. 2(A)). The second moment of this distribution is $u_q=x_0^2 q_0+x_1^2q_1+\cdots +x_M^2 q_M$, and that of the $MED$ is $u_{MED}=\theta_0(x_r+x_{r+1})-x_{r}x_{r+1}$ when $\theta_0\in [x_r,x_{r+1}]$. We are to show that $u_q\geq u_{MED}$.

We start with an auxiliary distribution on an auxiliary set of variables. Let $P_y=q_0+q_1+\cdots+q_r$ and $P_z=q_{r+1}+q_{r+2}+\cdots+q_M$ be a distribution over two 
variables, $y,z$ defined as (see Fig. 2(B)) 
\[
y=\frac{x_0q_0+x_1q_1+\cdots+x_rq_r}{q_0+q_1+\cdots+q_M}
\]
 
\[
z=\frac{x_{r+1}q_{r+1}+x_{r+2}q_{r+2}+\cdots+x_Mq_M}{q_{r+1}+q_{r+2}+\cdots+q_M}
\]
Clearly, $P_yy+P_zz=\theta_0$. This is a third distribution with the same expected value, and a second moment $u_A=\theta_0(y+z)-yz$. We complete the proof by showing that $u_q\geq u_A \geq u_{MED}$. The first inequality follows by adding the two cauchy schwarz inequalities: 
\[
P_yy^{2}\leq x_0^2 q_0+\cdots x_r^2 q_r\quad\&\quad P_zz^{2}\leq x_{r+1}^2 q_{r+1}+
\cdots x_M^2 q_M
\]
Noting that $y\leq x_{r}\leq \theta_0 \leq x_{r+1}\leq z$,
we obtain the second inequality: $u_A-u_P=(\theta_0-y)(z-x_{r+1})+(z-\theta_0)(x_r-y)\geq 0 $. This completes the proof. It is also a known optimization problem \cite{MED-1, MED-2}.

\begin{figure}
\includegraphics[scale=0.44]{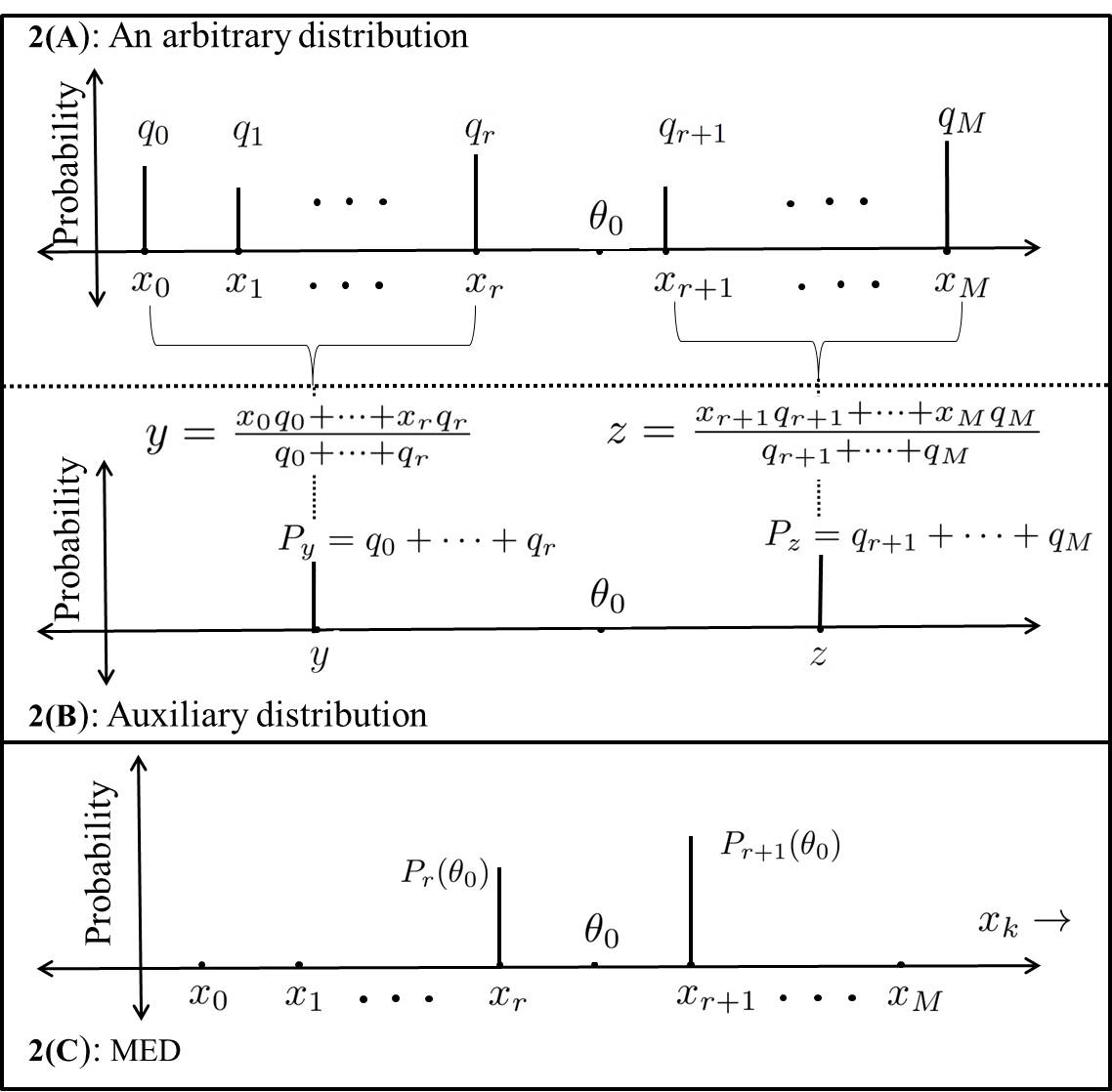}
\caption{(A.) An arbitrary distribution $\{q_0, q_1\cdots q_M\}$ on $\{x_0,x_1\cdots x_M\}$, such that the expected value $\theta_0=\sum_{k=0}^Mx_kq_k$ falls between $x_r$ and $x_{r+1}$ (B.) From this arbitrary distribution, an  auxiliary distribution, $\{P_y, P_z\}$ can be constructed with the auxiliary variables $\{y, z\}$. While this two-point distribution has the same expected value $\theta_0$, it has a lower error than the distribution $\{q_0,q_1\cdots q_M\}$. (C.) A plot of the minimum-error-distribution, $P^{MED}_k(\theta_0)$ against $x_k$. The probability is zero everywhere except at two points, $x_r$ and $x_{r+1}$. the RMSE of this distribution is lower than that of the auxiliary distribution, and hence lower than that of any arbitrary distribution with the same expected value.}. 
\end{figure}

\par It is easy to check that the RMSE or the error for the MED is 
\begin{equation}
\sigma^2_{MED}(\theta_0)=(\theta_0-x_{r})(x_{r+1}-\theta_0)
\end{equation}
which depends on the constraint $\theta_0 \in [x_r,x_{r+1}]$. 
This is maximum at the midpoint, $\theta_0= \frac{x_r+x_{r+1}}{2}$, with a value $\frac{x_{r+1}-x_r}{2}$.  However, it varies within the bin.
Therefore, in order to find a bound of RMSE or the variance corresponding to this distribution, one needs to consider the worst value it can take i.e. the maximal value of $\sigma^{MED}(\theta_0)$ for any $\theta_0$. For a specific bean, the maximum the variance can take is $\frac{x_{r+1}-x_r}{2}$. For any $\theta_0$, one therefore needs to maximize $\sigma^{MED}(\theta_0)$ over all bins, that is, over all $r=0,1,\cdots M-1$.  

\par To find the overall bound, one can first note that the maximum is always greater than its average value: $\frac{1}{M}\sum_r \frac{x_{r+1}-x_r}{2}=\frac{x_M-x_0}{2M}$. The equality holds when $x_0,\cdots x_M$ are uniformly spaced and therefore, this is the case when the error for $MED$ is the lowest: this value is $\frac{x_M-x_0}{2M}$. Furthermore, one can scale and translate $x_0$ and $x_M$ to $-1$ and $+1$ respectively, without loss of generality. The error of the $MED$ is then bounded by $\frac{1}{M}$. Note that this scaling, in effect, non-dimensionalizes the observable. Thus, the error is now a relative error, and hence dimensionless.

\par From (2), it may be noted that $\sigma^{MED}(\theta_0)$ is precisely zero at the end points of the interval: $\sigma_{MED}(x_r)=\sigma_{MED}(x_{r+1})=0$. However, these zero values do not imply sharp measurements or an absolute bound. 
For example, if $\theta_{est}=x_r$ is a value estimated from an actual measurement following the model, then, the range of the true value $\theta_0$ can only be inferred from condition $|\theta_0-x_r|\leq\sigma^{MED}(\theta_0)$. Clearly, the range of $\theta_0$ that satisfy this inequality is well within $[x_{r-1},x_{r+1}]$. This interval can be broken in to two neighboring intervals: $[x_{r-1},x_r] \cup [x_r,x_{r+1}]$.

$\sigma_{MED}(\theta_0)$ is different in these two neighboring intervals. In the former, it is $\sigma_{MED}^2(\theta_0)= (\theta_0-x_{r-1})(x_r-\theta_0)$. Therefore, within this interval, the range of $\theta_0$ satisfying the condition $|\theta_0-x_r|\leq\sigma^{MED}(\theta_0)$ is given by $(\theta_0-x_r)^2\leq(\theta_0-x_{r-1})(x_r-\theta_0)$. Or, equivalently,  $\theta_0 \in [\frac{x_r+x_{r-1}}{2},x_r]$. Similarly, the required range within the latter interval, $[x_{r},x_{r+1}]$, is $\theta_0 \in [x_r,\frac{x_r+x_{r+1}}{2}]$. This yields a total range of $\theta_0 \in [\frac{x_r+x_{r-1}}{2},\frac{x_r+x_{r+1}}{2}]$, that satisfies the above condition. This, indeed is $x_r \pm \frac{1}{M}$ when the $x_k$'s are all uniformly spaced, and $x_M$, $x_0$ are scaled to $\pm1$.

 \par This is our main result: We have explicitly constructed a distribution that minimizes the RMSE for a measurement model yielding $M+1$ discrete set of outcomes and found that the error bound scales as $1/M$. In what follows, we will analyze the implications and consequences of these findings.
 
 \subsection{Relation to Cramer-Rao bound}
 
 \par For a given random  variable $\{x_0 \cdots x_M\}$ and a given distribution $\{P_0(\theta_0)\cdots P_M(\theta_0)\}$, the Cramer-Rao bound sets a lower bound on the error in estimating the parameter $\theta_0$. This bound is expressed in terms of the score $S$, defined as $S_k=\frac{\partial}{\partial \theta_0}ln\left( P_k(\theta_0)\right)$. The square-error, $\sigma^2(\theta_0)$ is bounded below by the inverse of the Fisher information, $I(\theta_0)=\langle S^2\rangle $. 

\par One can note that, the Cramer-Rao bound does not provide an absolute bound on the attainable precision in a general estimation; it provides a bound for estimations using a particular distribution. This means, the bound for $\{P_0(\theta_0)\cdots P_M(\theta_0)\}$ may be surpassed by using a different distribution $\{P'_0(\theta_0)\cdots P'_M(\theta_0)\}$ for estimating the same parameter $\theta_0$. 

\par One can further note that, the Cramer-Rao bound does not guarantee attainability. Though it sets a lower bound on the error, an estimator which saturates this bound for the given distribution might not exist. The bound is saturated only for certain distributions. In this section we demonstrate that the MED saturates the inequality and it has the largest Fisher information among all those distributions that saturate the inequality. The score for MED is given by 
\begin{equation}
S_k(\theta_0)=\begin{cases}
\frac{1}{\theta_0-x_{k-1}} & x_{k-1}\leq \theta_0 \leq x_{k}\\
\frac{-1}{x_{k+1}-\theta_0} & x_{k}\leq \theta_0\leq x_{k+1}\\
0 & \theta_0>x_{k+1}  \quad or \quad \theta_0<x_{k-1}
\end{cases}
\end{equation}

\par Therefore, the Fisher information is $I_{MED}(\theta_0)=\langle S^2 \rangle= \frac{1}{(\theta_0-x_{r-1})(x_{r}-\theta_0)}$, for $x_{r-1}\leq \theta_0 \leq x_r$. Thus, the error bound is $\sigma \geq \sqrt{(\theta_0-x_{r-1})(x_{r}-\theta_0)}$. As seen in the main text, the RMSE of the MED is exactly equal to this expression.

\par It follows that the MED has the largest Fisher information among all those distributions that saturate the Cramer-Rao inequality. Let $\{P_0(\theta_0),P_1(\theta_0)\cdots P_M(\theta_0)\}$ be any distribution, and $I(\theta_0)$ be its Fisher information. If this distribution saturates the Cramer-Rao inequality for some estimator, the RMSE for that estimator will be $\frac{1}{\sqrt{I(\theta_0)}}$. Since the MED has the least RMSE, it follows that $I_{MED}(\theta_0)\geq I(\theta_0)$. Thus, the MED has the largest Fisher information, among all those distributions that saturate the Cramer-Rao inequality for some estimator.

 \subsection{Graphical Representation and Resource Count:} To analyze the physical consequences of MED and its bound, we introduce a simple graphical representation by comparing the error $\sigma(\theta_0)$ or equivalently, the second moment $u(\theta_0)=\sum_i x_k^2P_k(\theta_0) $ \cite{Variance-1} against $\theta_0$. 
After rescaling $x_M$ and $x_0$ to $\pm 1$, we have $\theta_0$ also bounded by $[-1,+1]$ and $u(\theta_0)$ lies in the range $[0,1]$. The graph is therefore confined to a rectangle (Fig.3). Furthermore, the absolute limiting parabola $u=\theta_0^2$ which sets the bound for zero error. Henceforth, this  parabola will be referred as the Zero Error Curve(ZEC). 

\begin{figure}
 \includegraphics[scale=0.32]{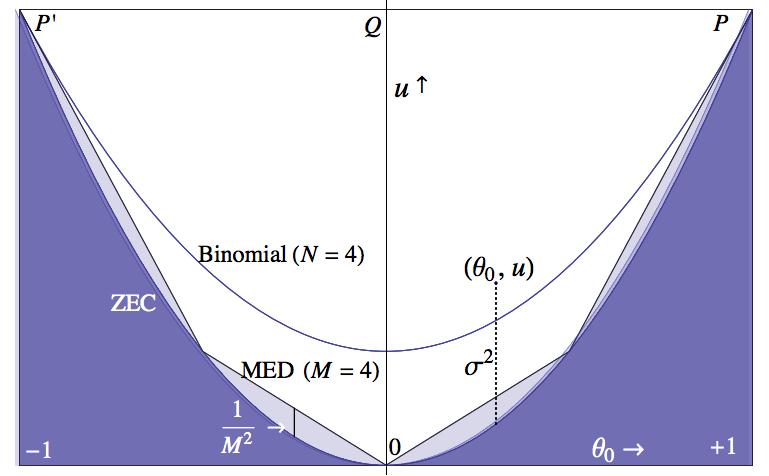}
\caption{A plot of the second moments against the expected value $\theta_0$: The limiting curve, or the ZEC (Zero error curve) is the parabola $u(\theta_0)=\theta_0^2$), such that the region below (shaded in dark) is physically inaccessible. Any measurement model is represented by a point $(\theta_0,u)$ with the RMSE given by the square root of its distance from the ZEC.
The polyline representing the second moment for the minimum-error-distribution, $u_{MED}$ is plotted for $M=4$. The region between this polyline and the ZEC (shaded in light) is inaccessible for any model with M+1 or lesser outcomes. The error-bound for the $MED$ corresponds to its optimal distance from the ZEC. The parabola for a binomial distribution (shown for $N = 4$, see text) is tangential to the corresponding $u_{MED}(\theta_0)$ at the endpoints, $\theta_0=\pm1$.}
\end{figure}
 
\par In order to visualize the MED, we note that the bound on the minimum error is achieved for equally spaced $x_k \in [-1,+1]$. This corresponds to $x_k=\frac{2k}{M}-1$, with $k=0,1,\cdots M$. The x-axis( Fig.3) can now be visualized as broken into $M$ bins, with the true value $\theta_0$ belonging to one such bin. Also, the second moment is 
\begin{equation}
u_{MED}(\theta_0)=\theta_0(x_r+x_{r+1})-x_rx_{r+1}.
\end{equation}
\par One can right away note that this is not a smooth function of $\theta_0$. Instead, it depends on the bin in which $\theta_0$ falls and is a straight line within a particular bin. Furthermore, the slope of the line depends on the specific bin and therefore on $\theta_0$, which makes the curve a polyline. Such a polyline for $M=4$ is shown in Fig. 3.  

\par Since these polylines correspond to the minimum error for a given $M$,  one can conclude that points which are below the $u_{MED}(\theta_0)$  and above the ZEC are physically inaccessible to any measurement model with $M+1$ distinct outcomes. For large $M$, the $u_{MED}(\theta_0)$ of-course approach the ZEC, as is expected physically.

\par The second moment of any arbitrary distribution on these $M+1$ variables lies above the polyline corresponding to $M$. For example, one can consider a binomial distribution with $M+1$ outcomes. This distribution can be constructed from a simple model with two possible measurement outcomes, $\pm1$, used to estimate the parameter $\theta_0$. The corresponding probabilities are $P_{\pm}=\frac{1\pm\theta_0}{2}$. For $N$ uncorrelated repetitions of this experiment, the resource count is $N$ together with $N+1$ possible outcomes, $\frac{\pm1 \pm1\cdots \pm1}{N}$ or equivalently, $x_k=\frac{2k}{N}-1$ (for $k=0,1\cdots N$). Here, $N=M$. The probabilities are binomial in $k$, with the second-moment  $u(\theta_0)=\theta_0^2+\frac{1-\theta_0^2}{N}$ representing a parabola (Fig. 3). 

\par Of course, this parabola lies completely above the $u_{MED}$ (corresponding to this $N$), which is tangential at the two end-points ($u'(\pm1) = u'_{MED}(\pm1)$). One can note that $u_{MED}$ for any lower number of outcomes cuts this parabola. But by definition, $u_{MED}$ is the limiting curve for any model with fixed outcomes. 

\par One can thereby conclude that for a binomial distribution with $N+1$ outcomes, there is no way to surpass its precision completely(at all $\theta_0$), using lower number of outcomes, for any arbitrary distributions. However, for same $N$, there can be several distributions with a tighter bound than the binomial and limited by the $MED$.

On the contrary, the precision of MED can be surpassed using uncorrelated, binomial distribution with larger number of outcomes. The number $M$ is therefore the only significant factor here for precision bound and not the actual distribution. This observation is consistent with our earlier result, that for a measurement with $M+1$ discrete outcomes, the precision is bounded by $\frac{1}{M}$, suggesting a possible characterization of the resource count using the number $M$ \cite{Ansatz-1}. We therefore propose an ansatz:

\par \textit{Ansatz:} M, the number of outcomes, is the resource count for a measurement. This implies, for a measurement model that uses $N$ resources, the error is absolutely bounded by $1/N$. 

In the rest of the paper, we will address the question of possibility of designing a physical experiment that achieves the bound of $MED$.

\section{Physical implications:} In order to attain the minimum-error, the system dynamics should result in the outcomes $\{x_0,x_1\cdots x_M\}$ being distributed following the $MED$(1).
 However, this distribution, unlike the binomial, is not smooth in $\theta_0$. This implies some amount of prior information of $\theta_0$ for choosing the appropriate system, probe and dynamics. The question then is how precise this prior information needs to be?
 One can right away observe that since the length of each bin is $\frac{2}{M}$, this amounts to demanding an apriori knowledge of $\theta_0$ with a precision $\frac{1}{M}$. Therefore, such a measurement procedure will not add any further value to the already known result. We illustrate this with an example.

\par \textit{A thought experiment:} Let us consider the well known example of estimating an unknown phase with a precision of $\frac{1}{N}$ using an $N$ particle entangled state. We shall demonstrate that this estimation is incomplete without attaining the MED in its final stage, thus demanding an apriori knowledge of the unknown phase.

In this case, a highly entangled input state $\psi_{in}=\frac{1}{\sqrt{2}}\{|\uparrow\rangle^{\otimes N}+|\downarrow\rangle^{\otimes N}\}$ evolves to $\psi_{out}=\frac{1}{\sqrt{2}}\{|\uparrow\rangle^{\otimes N}+e^{iN\phi_0}|\downarrow\rangle^{\otimes N}\}$, such that the phase $\phi_0$ is estimated from $\psi_{out}$, through an appropriate POVM. One choice of such a POVM is $|\psi_{in}\rangle\langle \psi_{in}|-|\tilde{\psi}_{in}\rangle\langle \tilde{\psi}_{in}|$, where $\tilde{\psi}_{in}=\frac{1}{\sqrt{2}}\{|\uparrow\rangle^{\otimes N}-|\downarrow\rangle^{\otimes N}\}$, is orthogonal to $\psi_{in}$. This yields an expectation value of $\theta_0=cosN\phi_0$ with an error of $sinN\phi_0$.  When repeated $\nu$ times, the error scales down to $\frac{sinN\phi_0}{\sqrt{\nu}}$. Therefore, a measurement result, $N\phi_{est}$, can be used to obtain an estimate of $\phi_0$ with a precision of $\frac{1}{N\sqrt{\nu}}$.

In this measurement, the last step is critical: obtaining $\phi_{est}$ from $N\phi_{est}$. For $N\phi_{est}$ in the range $[0,2\pi)$, there are $N$ possible values of $\phi_{est}$, given by $\{\frac{N\phi_{est}}{N},\frac{N\phi_{est}+2\pi}{N}\cdots \frac{N\phi_{est}+2\pi(N-1)}{N}\}$. This is a well known problem for unknown phase estimation, and several strategies are proposed to resolve the ambiguity \cite{Multiplicity-1, Multiplicity-2, Multiplicity-3, Multiplicity-4}.

Let S be such a resolution strategy, used as part of this measurement, to resolve among these $N$ values of $\phi_{est}$. The $N$ possible values of $\phi_{est}$ are spread out uniformly on the unit circle, with a gap of $\frac{2\pi}{N}$ between them. Therefore, the strategy S, is a measurement, that resolves these values with a precision of $\frac{\pi}{N}$. We shall now demonstrate that such a strategy leads to  $MED$ for the $\phi_{est}$ and therefore require an apriori knowledge of the phase.

As a resolution strategy, S associates each of the $N$ possible values of $\phi_{est}$ with a probability, such that the expected value is $\phi_0$.  Furthermore, to ensure that the strategy resolves the "correct" value, the RMSE of this distribution should not exceed $\frac{\pi}{N}$. Since these $N$ values are uniformly spread over an interval of $[0,2\pi)$, $\frac{\pi}{N}$ is indeed the minimum error that any distribution can attain. However, by uniqueness, the $MED$ is the only distribution that attains this value. Therefore, the strategy S involves attaining $MED$, over a set of $N$ variables, thereby demanding an apriori knowledge of the parameter. 

\section{Conclusions}  
  For a measurement with a fixed number of outcomes $\{x_0,x_1\cdots x_M\}$, i.e., for a fixed $M$ and for any probability mass-function $\{P_0(\theta_0), P_1(\theta_0)\cdots P_M(\theta_0)\}$, the attainable precision is shown to be bounded below by $\frac{x_M-x_0}{M}$. Therefore, the relative error scales inversely with $M$. Furthermore, the distribution which attains this bound (MED) is constructed explicitly, and it is found to be unique.

A simple ansatz is proposed equating the resource count to  the number of discrete outcomes, irrespective of the actual values $\{x_0, x_1 \cdots x_M\}$,or the probabilities. Therefore, in quantum metrology, one can now compare widely different measurement strategies, by counting only the possible outcomes the measurement model will yield. For equal number of probable outputs, one can conclude the resources used to be equal in otherwise differing experimental schemes. 

We observe that a physical implementation of MED demands an additional apriori information on the natural value $\theta_0$, in any strategy, making $M$ an under count. While this is known for phase estimation strategies, we show this result in general for any single parameter estimation. Since this is a theoretical limitation, we believe that this can be relevant for quantum foundations. 

In a quantum measurement, the resource count is the number of distinct eigen values of the observable. It depends only on the observable. In general, resource is maximum when the observable is totally non degenerate. That is, none of its eigen states are degenerate.
However, of practical relevance are systems of identical particles that are either symmetric or anti symmetric under particle exchange. Therefore, the observable has to satisfy an additional constraint- it should be invariant under all particle permutations. With this constraint, an observable is never totally non degenerate. Therefore, maximizing the resource count over this class of observables is nontrivial, and depends on the system. This is an open problem at the moment.

\par We acknowledge J. Dutta,  M. K. Harbola, J. Kolodynski, T. Moroder and V. Ravishankar for stimulating discussions and comments. This work was supported by the IIT-Kanpur Initiation Grant: IITK-PHY-20120111.

\end{document}